\documentclass[prb]{revtex4}
\usepackage{epsfig}
\usepackage{amsmath}
\begin{document}

\title{Persistent spin and charge currents and magnification effects in open ring conductors
subject to Rashba coupling}
\author{R. Citro and F. Romeo}
\affiliation{Dipartimento di Fisica ``E. R. Caianiello",
Universit{\`a} degli Studi di Salerno, and Unit{\`a} C.N.I.S.M.,
Via S. Allende, I-84081 Baronissi (SA), Italy }

\date{\today}

\begin{abstract}
We analyze the effect of Rashba spin-orbit coupling and of a local
tunnel barrier on the persistent spin and charge currents in a
one-dimensional conducting Aharonov-Bohm (AB) ring symmetrically
coupled to two leads. First, as an important consequence of the
spin-splitting, it is found that a persistent spin current can be
induced which is not simply proportional to the charge current.
Second, a magnification effect of the persistent spin current is
shown when one tunes the Fermi energy near the Fano-type
antiresonances of the total transmission coefficient governed by
the tunnel barrier strength. As an unambiguous signature of
spin-orbit coupling we also show the possibility to produce a
persistent {\it pure} spin current at the interference zeros of
the transmittance. This widens the possibilities of employing
mesoscopic conducting rings in phase-coherent spintronics
applications.
\end{abstract}


\maketitle

Recently enormous attention, from both experimental and
theoretical physics communities, has been devoted towards the
interplay of spin-orbit (SO) coupling and quantum interference
effects in confined semiconductor heterostructures. Such interplay
can be exploited as a mean to control and manipulate the spin
degree of freedom at mesoscopic scale useful for phase-coherent
spintronic applications.\cite{spintr_1,spintr_2} The major goal in
this field is the generation of spin-polarized currents and their
appropriate manipulation in a controllable environment. Since the
original proposal of the spin field effect transistor (spin FET)
by Datta and Das\cite{FET_datta}, many proposals have appeared
based on intrinsic spin splitting properties of semiconductors
associated with the Rashba spin orbit(SO)
coupling\cite{rashba_1,rashba_2}. This is a dominant mechanism
that has been proven to be a convenient mean of all-electrical
control of spin polarized current through additional gate
voltages.\cite{nitta_electric_contr}. In addition, suitable means
for controlling spin at mesoscopic scales are provided by quantum
interference effects in coherent ring conductors under the
influence of electromagnetic potentials, known as
Aharonov-Bohm(AB)\cite{aharonov_bohm} and
Aharonov-Casher(AC)\cite{aharonov_casher} effect. This possibility
has driven a wide interest in spin-dependent transmission
properties of mesoscopic AB and AC rings that have been studied
under various aspects
\cite{buttiker_ring,loss_ring,stern_ring,aronov_ring,qian_ring,nitta_ring_so,frustaglia_ring_so,sigrist_as,foldi_05}.
In particular, an extended literature is devoted to the study of
persistent currents, from both the experimental and theoretical
point of view, in closed and open quantum rings. In this context,
only few papers have appeared on the analysis of persistent
currents in rings where the Rashba effect is likely to be
important\cite{governale_ring,zhang_pcso,gritsev_04,sun_06,chang_06}.
In the early work of Splettstoesser et al.\cite{governale_ring}
the analysis of the persistent current in a ballistic mesoscopic
ring with Rashba SO coupling and in absence of current leads has
been employed to extract the strength of Rashba spin splitting.
Unambiguous signatures of SO have also been found in
Ref.[\onlinecite{zhang_pcso}] where the persistent spin current
has been analyzed in a quantum ring with multiple arms and in
presence of current leads. Here it is found that the SO coupling
can increase or decrease the total persistent currents and change
its direction. In this paper we focus on the persistent spin and
charge currents in a one-dimensional AB ring connected to external
leads, interrupted by a tunnel barrier in the lower arm and
subject to Rashba spin-orbit interaction\cite{note_dresselhaus}.
First, we revisit the subject of the spin-splitting effect on the
persistent currents as a function of the Rashba coupling strength,
then we extract distinct effects due to the presence of the tunnel
barrier in one of the arms which have not been considered in
earlier works. In particular, we show a magnification effect of
the persistent spin current when the Fermi energy is tuned near
the antiresonances of the total transmission coefficient caused by
the presence of the tunnel barrier. Such effect is similar to the
one discussed in Ref.[\onlinecite{yi_double_ring}] where the
magnification effect on persistent currents is discussed for an
open Arhonov-Bohm double-ring structure in absence of SO coupling.
Jayannavar et al.\cite{deo_95} have also reported magnification
features of the persistent currents near the conductance
antiresonances in a ring with rotational symmetry breaking due to
unequal arm lengths but in absence of magnetic flux. As far as we
know, magnification effects on spin-currents in presence of Rashba
SO interaction have not been reported yet.  An important feature
is represented by the possibility of having a sizeable {\it pure}
spin current (in correspondence of zero charge current) at the
interference zeros of the transmittance.
An analysis of the persistent currents as a function of the
effective flux induced by the spin-orbit interaction is also
presented.



\begin{figure}[htbp]
\centering
\includegraphics[width=8cm]{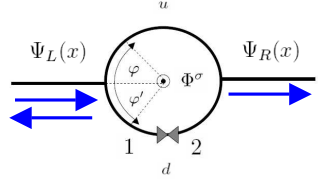}\\
\caption[Fig.0]{One dimensional ring in presence of current leads
and subject to Rashba spin-orbit interaction.}\label{fig:device}
\end{figure}

 The system under study is depicted in
Fig.\ref{fig:device}. In one dimensional rings on semiconductor
structure, an effective Rashba electric field results from the
asymmetric confinement along the direction ($k$) perpendicular to
the plane of the ring\cite{rashba_1,rashba_2}.
The Hamiltonian describing the Rashba SO coupling is the
following:
\begin{equation}
\hat{H}_{SO}=\frac \alpha \hbar
(\hat{\overrightarrow{\sigma}}\times \hat{\overrightarrow{p}})_k,
\end{equation}
where $\hbar/2 \hat{\overrightarrow{\sigma}}$ is the spin operator
expressed in terms of the Pauli spin matrices,
$\hat{\overrightarrow{\sigma}}=(\sigma_i,\sigma_j,\sigma_k)$  and
$\alpha$ is the SO coupling (SOC) associated to the effective
electric field along the $k$ direction. The total Hamiltonian of a
moving electron in presence of SOC can be found in
Ref.[\onlinecite{meijer_ham}]. In the case of a one-dimensional
ring an additional confining potential (e.g. of harmonic type)
must be added in order to force the electron wave function to be
localized on the ring. When only the lowest radial mode is taken
into account, the resulting effective one dimensional Hamiltonian
in a dimensionless form\cite{frustaglia_ring_so,molnar_ring_so}
can be written as:
\begin{equation}
\label{eq:ham_1d} {\hat H}=\frac{2m^\star
R^2}{\hbar^2}\hat{H}_{1D}=\left( -i \frac{\partial}{\partial
\varphi}+\frac{\beta}{2}\sigma_r-\frac{\Phi_{AB}}{\phi_0}\right)^2,
\end{equation}
where $m^\star$ is the effective mass of the carrier,
$\beta=2\alpha m^\star/\hbar^2$ is the dimensionless SOC,
$\sigma_r=\cos \varphi \sigma_x+\sin \varphi \sigma_y$, and
additional constants have been dropped, $\Phi_{AB}$ is the
Bohm-Aharonov flux and $\phi_0$ is the quantum flux $\phi_0=h c
/e$. The parameter $\alpha$ represents the average electric field
along the $k$ direction and is assumed to be a tunable quantity.
For an InGaAs-based two-dimensional electron gas, $\alpha$ can be
controlled by a gate voltage with typical values in the range
(0.5$-$2.0)$\times 10^{-11}$eVm\cite{param1,param2}. The tunnel
barrier localized in the lower arm of the ring is modelled by a
delta potential, $v\delta(\varphi'+\frac{\pi}{2})$, where $v$ is
the dimensionless tunnel barrier strength $v=2m^\star R^2
V/\hbar^2$, and $\varphi'=-\varphi$. The local tunnel barrier can
be experimentally realized by a quantum point
contact\cite{ring_exp} and its strength controlled by a so-called
split-gate voltage. As outlined in the Appendix of
Ref.[\onlinecite{molnar_ring_so}] when $v$ is zero, one can solve
the eigenvalue problem in a straightforward manner and the energy
eigenvalues are:
\begin{equation}
E^\sigma_n=(n-\Phi_{AC}^\sigma/2\pi-\Phi_{AB}/2\pi)^2,
\end{equation}
where $\sigma=\pm$, $\Phi_{AC}^\sigma$ is the so-called
Aharonov-Casher phase\cite{aharonov_casher}
$\Phi_{AC}^\sigma=-\pi(1-\sigma \sqrt{\beta^2+1})$ At fixed
energy, the dispersion relation yields the quantum numbers
$n_\lambda^\sigma (E)=\lambda \sqrt{E}+\Phi^\sigma/2\pi$, where we
have introduced $\Phi^\sigma=\Phi_{AC}^\sigma+\Phi_{AB}$, and the
index $\lambda=\pm$ refers to right/left movers, respectively. The
unnormalized eigenvectors have the general
form\cite{frustaglia_ring_so,molnar_ring_so} $
\Psi^\sigma_n(\varphi)=e^{in\varphi}\chi^\sigma(\varphi)$, where
$n\in$Z is the orbital quantum number.  It should be noted that
the spinors $\chi^\sigma(\varphi)$ are generally not aligned with
the Rashba electric field, but they form a tilt angle given by
$\tan \theta=-\beta$ relative to the $k$ direction and can
expressed in terms of the eigenvectors of the Pauli matrix
$\sigma_k$\cite{molnar_ring_so}.
For our purposes, we set up a scattering problem and calculate the
transmission coefficient, following the method of quantum
waveguide transport on networks\cite{xia_waveguide,deo_94}. One
main problem is the boundary conditions at the intersection with
the external leads and at the tunnel barrier. In this case it is
appropriate to apply the spin-dependent version of the Griffith
boundary's condition\cite{griffith_boundary}. These state that (i)
the wave function must be continuous and (ii) the spin density
must be conserved. The same conditions apply at the location of
the tunnel barrier in the
lower arm\cite{citro_06}. 

We assume that when an electron moves along the upper arm in the
clockwise direction from  $\varphi=0$ (see Fig.\ref{fig:device}),
it acquires a phase $\Phi^\sigma/2$ at the output intersection
$\varphi=\pi$, whereas the electron acquires a phase
$-\Phi^\sigma/4$ in the counterclockwise direction along the lower
arm when moving from $\varphi'=0(\pi/2)$ to $\varphi'=\pi/2(\pi)$.
Therefore the total phase is $\Phi^\sigma$ when the electron goes
through the loop. The wave functions in the upper($u$) and
lower($d$) arm of the ring can be written as:
\begin{eqnarray}
\Psi_{u}(\varphi)=\sum_{\sigma=\pm,\lambda=\pm}
c_{u,\sigma}^\lambda e^{i n_\lambda^\sigma \varphi}
\chi^\sigma(\varphi), \nonumber \\
\Psi_{d\alpha}(\varphi')=\sum_{\sigma=\pm,\lambda=\pm}
c_{d\alpha,\sigma}^\lambda e^{-in_\lambda^\sigma
\varphi'}\chi^\sigma(\varphi'),
\end{eqnarray}
where the index $d\alpha=d1,d2$ denotes the wave function in the
two-halves of the lower branch and $n_\lambda^\sigma=\lambda k
R+\Phi^\sigma/2\pi$. The wave function of the electron incident
from the left lead in the left and right electrodes can be
expanded as:
\begin{equation}
\Psi_L(x)=\Psi_i+(r_\uparrow,r_\downarrow)^Te^{-ik x},\mbox{
}\Psi_R(x)=(t_\uparrow,t_\downarrow)^T e^{ik x},
\end{equation}
where $x=R \varphi$, $r_\sigma$ and $t_\sigma$ are the
spin-dependent reflection and transmission coefficient, $\Psi_i$
is the wave function of the injected electron $\Psi_{i}=e^{ik
x}\chi^\sigma(0)$. For an incident electron from the right lead an
analogous expansion is possible. This enables us to formulate the
scattering matrix equation as $\hat{o}=\hat{S}\hat{i}$, where
$\hat{o},\hat{i}$ stand for outgoing and incoming wave
coefficients. By applying the boundary conditions at the
junctions\cite{citro_06} and the conservation of the currents, we
have a set of equations that can be solved with respect to the
transmission coefficients $t_\sigma(kL,\Phi^\sigma,z)$, where we
have used $z=v/k$ and $L=\pi R$:
\begin{eqnarray}
\label{eq:trans} t_\sigma(kL,\Phi^\sigma,z)=\frac{8\,\sin
(\frac{kL}{2})\left( -4\,\cos (\frac{kL}{2})\,\cos
(\frac{\Phi^{\sigma}}{2} ) + z\,\sin
(\frac{kL}{2})\,e^{i\frac{\Phi^{\sigma}}{2}}\right)}
    {4\,z\,\cos (\frac{kL}{2}) - 2\,\left( 5\,i + 2\,z \right) \,\cos (2kL) +
    i\,\left( 2 + 8\,\cos (\Phi^{\sigma} ) - 2\,z\,\sin (\frac{kL}{2}) + \left( 8\,i + 5\,z \right) \,\sin (2kL) \right)}.
\end{eqnarray}
The transmission probability (or transmittance) in the spin
channel $\sigma$ is given by $T_\sigma=t_\sigma^\star t_\sigma$
and is related to conductance via the well-known
Landauer-B\"{u}ttiker formula\cite{landauer_cond}. $T_\sigma(kL,
\Phi^\sigma,z)$ is periodic in $kL$ with period $2\pi$ and in
$\Phi^\sigma$ with period $2\pi$. Therefore in the following we
only consider the region $0\le kL \le 2\pi$ and $0\le \Phi^\sigma
\le 2\pi$. A remarkable feature is that the transmittance presents
both resonances and antiresonances. The antiresonances become
asymmetric or of Fano-type\cite{fano} in the presence of a finite
tunnel barrier strength\cite{citro_06}. When $z=0$ only symmetric
antiresonances are possible, while only resonances exist in the
single ring without a barrier and in absence of electromagnetic
fluxes.

 We now examine the
current flows in the AB-AC ring. As electrons carry spin besides
charge, their motion gives rise to a spin current other than a
charge current. The difference of charge current carried by
spin-up and spin-down electrons is identified with the spin
current\cite{governale_ring}, $I_s=\sum_{\sigma=\pm} \sigma
I_\sigma$. In the presence of spin-orbit interaction the spin
currents are not simply proportional to the charge currents. As we
will show a possibility could emerge in which the spin current
becomes {\it pure}, i.e. when the charge current, $I_c=\sum_\sigma
I_\sigma$, is exactly zero. The spin currents in the upper and
lower arm are generally different by various symmetry breaking. In
the case under investigation both time-reversal symmetry and
spin-reflection symmetry are broken via the Aharonov-Bohm and
Aharonov-Casher effect, respectively, while the tunnel barrier
breaks the rotational symmetry. This is responsible for the
persistent charge and spin current in the ring. When the current
in one arm is larger than $T_\sigma$ , the current in the other
arm has to be negative to conserve the total current at the
junction with the external leads. One can view such a negative
current as a circulating current in the loop and define it as a
persistent current.\cite{deo_95} The probability current in the
upper arm is given by
$I^u_\sigma=(|c^{+}_{u,\sigma}|^2-|c^{-}_{u,\sigma}|^2)$ and can
be written as:
\begin{equation}
\label{eq:current} I^u_\sigma=\frac{ T_\sigma}{ 2}\left(1-2
\frac{\tan \pi \Phi^\sigma}{\tan (k L )}\right)-F(k L
,\Phi^\sigma,z),
\end{equation}
where $F(k L,\Phi^\sigma,z)$ is complicated function of the
energy, the effective flux and the tunnel barrier. With the above
definition, the persistent current with spin $\sigma$ is given by
$I_{p\sigma}=(T_\sigma-I^u_\sigma)$, when $I^u_\sigma>T_\sigma$
and $I_{p\sigma}=-I^u_\sigma$ when $I^u_\sigma$ is negative.  A
similar definition holds when we consider the probability current
in the lower arm $I^d_\sigma$. After having identified the
wave-vectors intervals wherein either $I^u_\sigma$ or $I^d_\sigma$
flow in the negative direction, by their magnitudes we have
calculated the persistent currents per spin $\sigma$ and the
persistent charge and spin currents by: ${\cal I}_s=\sum_\sigma
\sigma I_{p\sigma}$ and ${\cal I}_c=\sum_\sigma I_{p\sigma}$.

\begin{figure}[htbp]
\centering
\includegraphics[height=6cm,width=9cm]{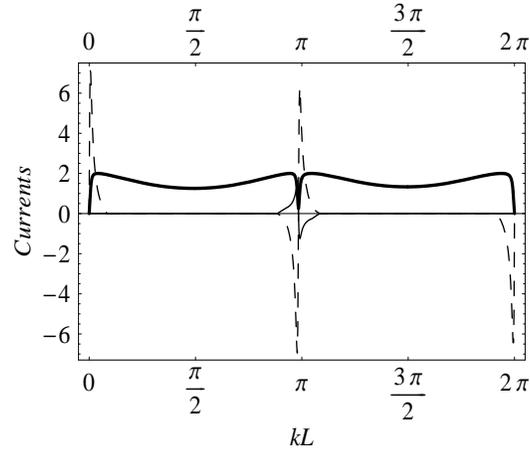}\\
\caption[Fig.3]{Persistent currents $\mathcal{I}_{c}$ (thin line)
and $\mathcal{I}_{s}$ (dashed line) in dimensionless units (of $e
v_F$ and $\hbar/2 v_F$, respectively) plotted as a function of
$kL$ with $\Phi_{AB}/(2\pi)=0.5$, $\beta=\sqrt{3}+ 0.1$, $z=0.1$.
The persistent currents are magnified in vicinity of a Fano-like
anti-resonance in the normalized transmittance (thick
line).}\label{fig:curr_energy}
\end{figure}

\begin{figure}[htbp]
\centering
\includegraphics[width=8cm]{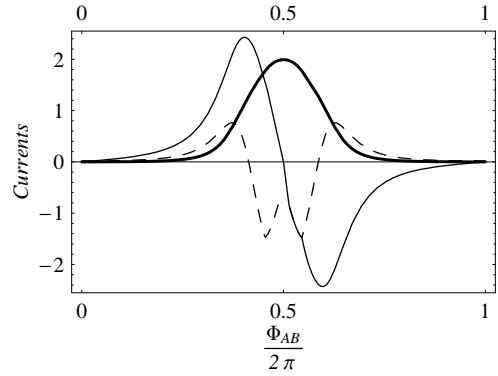}\\
\caption[Fig.7]{Persistent currents $\mathcal{I}_{c}$ (thin line)
and $\mathcal{I}_{s}$ (dashed line) in dimensionless units (see
text) vs $\Phi_{AB}/(2\pi)$ with $kL=2\pi+0.1$, $\beta=1.8$,
$z=0.5$. A pure spin persistent currents is obtained for
half-integers values of the Aharonov-Bohm flux. The thick line
represents the transmittance.} \label{fig:curr_flux}
\end{figure}

\begin{figure}[htbp]
\centering
\includegraphics[width=8cm]{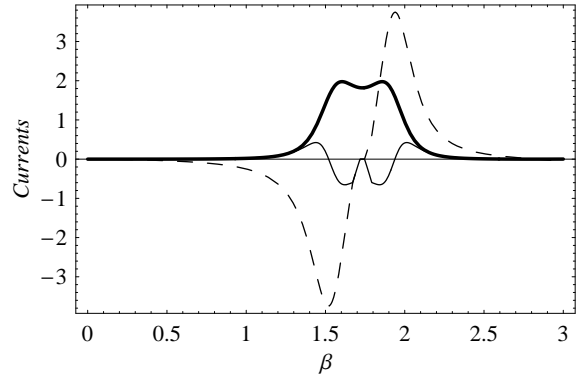}\\
\caption[Fig.8]{Persistent currents $\mathcal{I}_{c}$ (thin line)
and $\mathcal{I}_{s}$ (dashed line) in dimensionless units (see
text)plotted as a function of $\beta$ with $kL=2\pi+0.1$,
$\Phi_{AB}/(2\pi)=0.5$, $z=0.5$. The thick line represents the
transmittance.} \label{fig:curr_beta}
\end{figure}

\begin{figure}[htbp]
\centering
\includegraphics[width=8cm]{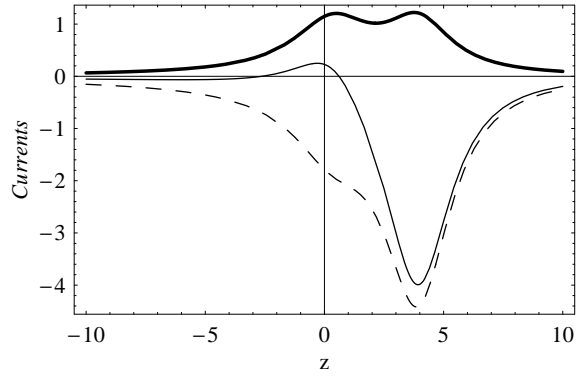}\\
\caption[Fig.13]{Persistent currents $\mathcal{I}_{c}$ (thin line)
and $\mathcal{I}_{s}$ (dashed line) in dimensionless units (see
text) plotted as a function of $z$ with $\Phi_{AB}/(2\pi)=0.45$,
$kL=2\pi+0.08$, $\beta=1.5$. The normalized transmittance (thick
line) is also shown.} \label{fig:curr_zeta}
\end{figure}

The transmittance and the persistent spin and charge currents in
dimensionless units (respectively of $\hbar v_F/2$ and $e v_F$ ,
$v_F$ being the Fermi velocity) are shown as a function of $kL$
($k$ near $k_F$) in Fig.\ref{fig:curr_energy} for
$\Phi_{AB}=\pi,\beta=1.83,z=0.1$. A remarkable feature is that in
correspondence of the Fano resonance at $kL=\pi$ the amplitude of
the spin-current is magnified and remarkably the spin current
becomes {\it pure} at the interference zeros, $kL=0,\pi$.  Let us
note that the amplitude of the persistent current is proportional
to the slope of the Fano resonance. Such a slope diverges when
$kL$ approaches the singular points and so does the persistent
current. The Fano-type resonances are present only when the tunnel
barrier strength is non-zero, as discussed above. We also notice
that the persistent currents change sign when crossing the energy
or the wave vector at the antiresonance. In fact such
antiresonance is characterized by an asymmetric pole structure in
the transmission amplitude. This behavior is similar to one
observed for the persistent charge currents in an open ring with
incommensurate arm lengths in absence of electromagnetic
fluxes\cite{deo_95}.
The origin instead of a pure spin current near $kL=0,2\pi$ stems
for the interference effects at the junctions in the presence of a
SO interaction that induces finite transmission probability in the
spin channel opposite to the incident spin
orientation\cite{bulgakov}. The divergent feature of the current
near $k L=0,2\pi$ stems from the first term in (\ref{eq:current})
while the current divergence near $k L=\pi$ stems for the second
term in (\ref{eq:current}) and has a non-trivial dependence on the
tunnel barrier. The persistent currents as a function of the
Aharonov-Bohm flux are reported in Fig.\ref{fig:curr_flux} for
$\beta=1.8,z=0.1,kL=2\pi$. Close to the maximum of the
transmittance at $\Phi_{AB}=\pi$ a pure spin current is detected.
Nearby two values of the Aharonov-Bohm flux ($0.8 \pi$ and $1.2
\pi$) correspond to a pure charge current. The persistent charge
and spin currents are shown in Fig.\ref{fig:curr_beta} as a
function of the SO coupling for
$\Phi_{AB}/2\pi=0.49,kL=2\pi,z=0.5$. For the values of the
parameters chosen, the spin current is magnified by varying the SO
coupling strength at the points where the transmittance has
maxima. Further the persistent spin current oscillates between
positive and negative values as the intensity of the SO coupling
increases. These features further indicate that the directions of
the persistent currents depend on the intensity of the SO
coupling, and that it can increase or decrease the total
persistent current. These findings are in agreement with those in
Ref.[\onlinecite{zhang_pcso}] where persistent spin currents in a
multiple arms ring were discussed. As a function of the tunnel
barrier strength $z$ the persistent spin current shows a minimum
without sign change. This implies that $z$ can be varied to
maximize the spin currents. This is shown in
Fig.\ref{fig:curr_zeta} where the parameters are
$\beta=1.5,\Phi_{AB}/2\pi=0.45$. Finally we have verified that
finite temperature effects do not lead to cancellations of the
persistent current features described above, apart from a slight
renormalization of the currents magnitude up to temperatures of
the order 100$mK$\cite{citro_06}at which real devices are working.
%
In conclusion, we have analyzed the properties of the persistent
spin and charge currents in an open quantum ring  subject to the
Rashba spin-orbit interaction in presence of an external magnetic
flux and a tunnel barrier in the lower arm. We have discussed a
magnification effect of the persistent spin currents in
association with the Fano resonances of the transmission
coefficient, depending on the magnitude of the tunnel barrier
strength $z$.
We have also shown that persistent {\it pure} spin currents can
arise which stem for the time-reversal symmetry breaking and the
spin-reversal symmetry breaking due to the total effective flux
enclosed in the ring. Finally, we have shown that the directions
of the persistent currents depend on the intensity of spin-orbit
coupling and the tunnel barrier strength that can increase or
decrease the persistent currents. The different dependencies of
the persistent charge and spin currents are a {\it unique}
signature of the spin-orbit coupling affecting the electronic
structure of the ring that can be exploited in experiments.
Indeed, the possibility to measure spin persistent currents in
open rings is within reach with today's technology for experiments
in semiconductor heterostructures, e.g. InGaAs-based
2DEG\cite{param1,ring_exp_AC}. Indeed spin interference effects in
Rashba-gate-controlled ring with a quantum point contact inserted
have recently been reported\cite{ring_exp} and could be further
investigated to reproduced the spin current magnification effects
discussed here.
\begin{acknowledgments}
We thank Prof. Maria Marinaro for very useful suggestions and
lightening discussions.
\end{acknowledgments}




\begin{thebibliography}{99}
\bibitem{note_dresselhaus} We have neglected Dresselhaus spin-orbit coupling having in mind, e.g.
InAs semiconductor where the Rashba interaction dominates.
\end{thebibliography}
\end{document}